\documentclass[sigconf, nonacm]{acmart}

\AtBeginDocument{%
  \providecommand\BibTeX{{%
    \normalfont B\kern-0.5em{\scshape i\kern-0.25em b}\kern-0.8em\TeX}}}

\copyrightyear{2025}
\acmYear{2025}
\setcopyright{cc}
\setcctype{by}
\acmConference[SC '25]{The International Conference for High Performance Computing, Networking, Storage and Analysis}{November 16--21, 2025}{St Louis, MO, USA}
\acmBooktitle{The International Conference for High Performance Computing, Networking, Storage and Analysis (SC '25), November 16--21, 2025, St Louis, MO, USA}
\acmDOI{10.1145/3712285.3759858}
\acmISBN{979-8-4007-1466-5/2025/11}

\usepackage{algorithmic}
\usepackage{textcomp}
\usepackage{xcolor, url}
\usepackage{balance}
\usepackage{tikz}
\usepackage[detect-all,group-separator={,}]{siunitx}
\usepackage{{booktabs}}
\usepackage{comment}
\usepackage{float}
\usepackage{multirow}
\usepackage{subcaption}
\usepackage{printlen}
\usepackage[font=small,skip=3pt]{caption}

\newcommand{\job}{job }

\newcommand{\jobs}{jobs }

\newcommand{\machine}{machine }
\newcommand{\machinenospace}{machine}
\newcommand{\machines}{machines }

\newcommand{\sysname}{\textit{green-ACCESS}}
\newcommand{\Sysname}{\textit{Green-ACCESS}}
\newcommand{\circled}[1]{\tikz[baseline=(char.base)]{
            \node[shape=circle,draw,inner sep=1pt] (char) {#1};}}

\begin{document}

\title{Core Hours and Carbon Credits: Incentivizing Sustainability in HPC}
\author[A. Kamatar]{Alok Kamatar}
\affiliation{%
  \institution{University of Chicago}
  \city{Chicago} 
  \country{USA} 
}
\email{alokvk2@uchicago.edu}

\author[M. Gonthier]{Maxime Gonthier}
\affiliation{%
  \institution{University of Chicago}
  \city{Chicago} 
  \country{USA} 
}
\email{mgonthier@uchicago.edu}

\author[V. Hayot-Sasson]{Valerie Hayot-Sasson}
\affiliation{%
  \institution{University of Chicago}
  \city{Chicago} 
  \country{USA} 
}
\email{vhayot@uchicago.edu}

\author[A. Bauer]{André Bauer}
\affiliation{%
  \institution{Illinois Institute of Technology}
  \city{Chicago} 
  \country{USA} 
}
\email{abauer7@illinoistech.edu}

\author[M. Copik]{Marcin Copik}
\affiliation{%
  \institution{ETH Zürich}
  \city{Zürich} 
  \country{Switzerland} 
}
\email{marcin.copik@inf.ethz.ch}

\author[R. Castro Fernandez]{Raul Castro Fernandez}
\affiliation{%
  \institution{University of Chicago}
  \city{Chicago} 
  \country{USA} 
}
\email{raulcf@uchicago.edu}

\author[T. Hoefler]{Torsten Hoefler}
\affiliation{%
  \institution{ETH Zürich \& Swiss National Supercomputing Center (CSCS)}
  \city{Zürich} 
  \country{Switzerland} 
}
\email{htor@inf.ethz.ch}

\author[K. Chard]{Kyle Chard}
\affiliation{%
  \institution{University of Chicago \& Argonne National Laboratory}
  \city{Chicago} 
  \country{USA} 
}
\email{chard@uchicago.edu}

\author[I. Foster]{Ian Foster}
\affiliation{%
  \institution{University of Chicago \& Argonne National Laboratory}
  \city{Chicago} 
  \country{USA} 
}
\email{foster@uchicago.edu}

\renewcommand{\shorttitle}{Core Hours and Carbon Credits}
\renewcommand{\shortauthors}{Kamatar et al.}

\begin{abstract}
Efforts to reduce the environmental impact of HPC often focus on resource providers, but choices made by users, e.g., concerning where to run, can be equally consequential.
Here we present evidence that new accounting methods that charge users for energy used can incentivize significantly more efficient behavior.
We first survey 300 HPC users and find that fewer than 30\% are aware of their energy consumption, and that energy efficiency is a low priority concern.
We then propose two new multi-resource accounting methods that charge for computations based on their energy consumption or carbon footprint, respectively.
Finally, we conduct both simulation studies and a user study to evaluate the impact of these two methods on user behavior.
We find that while only providing users feedback on their energy use had no impact on their behavior, associating energy with cost incentivized users to select more efficient resources, and use 40\% less energy.
\end{abstract}



\maketitle

\section{Introduction}

HPC sustainability is a shared responsibility between infrastructure \textit{providers} (who buy servers, source electricity, build and cool data centers, etc.) and \textit{consumers} (who write software and decide what, where, and when to compute~\cite{amazon-sustainability-model}). These roles should be complementary; however, they typically operate disjointly. 
Providers tout efficient facilities and investments in renewable/low-carbon energy~\cite{amazon-sustainability, google-sustainability, lumi2024sustainability} but are constrained by the amount, time, and location of consumer demand. 
Consumers are told that they can reduce environmental impact by using specific hardware or software, or by improving utilization, but lack information to accurately account for energy or carbon use. 
And while many tools exist to improve energy efficiency, these tools must first be adopted by users~\cite{geopm, app-aware-geopm, you2023zeus, chung2024perseus} or 
require cooperation between users and providers~\cite{sukprasert2023quantifying, wiesner2021letswait, fan-zcc, google-dc-shifting}. 
We argue and demonstrate that the key to achieving this cooperation is to charge users for energy/carbon used to incentivize more efficient behavior. 

We first survey more than 300 HPC users to assess their knowledge, attitudes, and behaviors with respect to energy consumption. 
This pioneering survey 
of HPC users on energy awareness and sustainability 
reveals \emph{a prevailing lack of awareness of energy use, lack of action to reduce energy, disregard for efficiency or sustainability rankings, and a low priority on energy efficiency.}  

To address these issues, we investigate
\emph{adopting an application's environmental impact as the measure of its resource usage}. 
We formulate two versions of impact-based accounting: \textbf{Energy-Based Accounting (EBA)} which charges jobs based on energy used, and \textbf{Carbon-Based Accounting (CBA)}, which charges jobs based on estimated carbon footprint.
Thus, a user might be allocated 10~kg carbon emissions (kgCO\textsubscript{2}e), rather than 100 node-hours; be able to estimate the kgCO\textsubscript{2}e required for a computation on different machines; and track kgCO\textsubscript{2}e rather than node-hours.
We envisage using EBA and CBA as the basis for fungible multi-resource HPC allocations where users choose between one of several machines to which they have access~\cite{boerner-access}. 
Previous approaches 
for incorporating energy cost into cloud  prices~\cite{hinz2018cost,kurpicz2016evape,aldossary2018energycost,narayan2014metering} are not directly applicable to HPC:
they aim to set cloud resource prices to recoup costs, not a primary concern in HPC allocations,
and do not account for carbon emissions.
We hypothesize that the adoption of such accounting schemes will incentivize users to develop more energy-efficient applications and select resources that are more energy efficient, allowing them to compute more for the same cost.

To enable experimentation with impact-based accounting 
across heterogeneous machines, we develop \Sysname{}, which leverages a Function-as-a-Service (FaaS) interface on top of HPC infrastructure~\cite{chard-funcx}.
This system makes energy consumption transparent, seamlessly guides users to more efficient machines, and incentivizes sustainable use of computing resources. 
With this system, we illustrate trade-offs between energy, time, and carbon-footprint, where, for instance charging based on peak-performance leads to a cheaper price for a machine that uses twice the energy.
Finally, we design a novel user-study using a web-based game to ascertain how energy information and impact-based accounting affect user behavior. We find that \emph{showing users their energy-use has no impact on behavior, but linking price with energy can encourage users to run on more efficient devices.}


The contributions of this paper are:
\begin{itemize}
    \item The first large-scale survey of 300+ HPC users focused on their awareness of energy and carbon use. We release the aggregate data to the community.
    \item Energy- and Carbon-Based Accounting (EBA and CBA) mechanisms for charging for computing based on environmental impact.
    \item A FaaS platform that implements EBA and CBA on diverse CPUs and GPUs. We open source the platform to be used as a plug-in on top of Globus Compute.
    \item A user study in which users interact with impact-based accounting via a web-based simulation game. 
\end{itemize}

\section{Survey}
 
We first examine the prevailing attitudes of HPC users towards sustainable computing practices. Specifically, we developed a survey to understand 
users' awareness 
of existing sustainability metrics and techniques. 

\subsection{Survey Design}
We employed Qualtrics, a platform for assembling and managing online surveys. We defined 33 questions that 
examine a participant's familiarity with their energy consumption and other sustainability-related questions. 
Our target participant was anyone who submits jobs or develops code for HPC machines. 
As such, we distributed the survey to HPC user groups, facilities, science collaborations known for use of HPC, and user groups of popular HPC tools.

\subsection{Results}
\label{sec:survey-results}

We received 316 responses, of which 192 completed 90\% of the survey or more. 
166 respondents were located in Europe, 104 in North America, 4 in Oceania, 4 in China, and 38 declined to share their location.
73 respondents were graduate students, 97 were early career researchers/engineers, and 99 were senior researchers/engineers.
We ask respondents to self-identify as HPC users and report the amount they used HPC. There was no significant difference in responses based on career stage or node-hours used.

\begin{figure}
    \centering
    \includegraphics[width=0.9\linewidth]{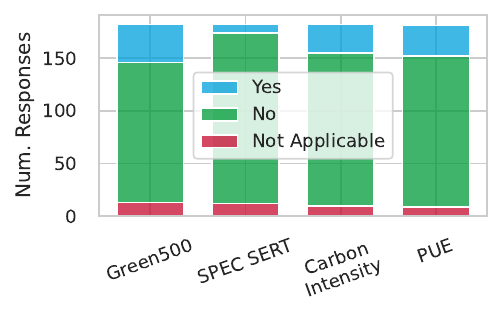}
    \Description[A stacked bar graph showing the number of respondents aware of sustainability metrics.]{The stacked bars show that most people are not aware of how their HPC machine performs on these metrics. The bar farthest to the right shows > 100 people are not aware of Green500.}
    \caption{Responses to question ``Are you aware of how the HPC resources you use perform on the following sustainability metrics?''}
    \label{fig:survey-tools}
\end{figure}

\begin{figure}
    \centering
    \includegraphics[width=0.86\linewidth]{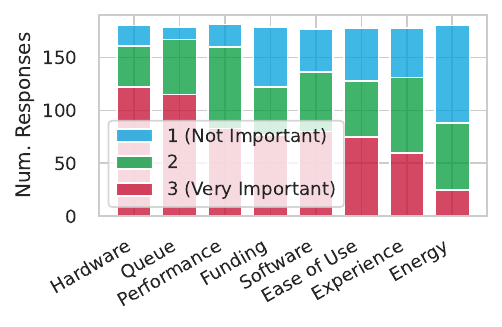}
    \Description[A stacked bar graph showing users priority when selecting resources.]{The stacked bars show that less than 50\% of people consider energy-efficiency as somewhat or very important.}
    \caption{Energy efficiency is among the least important factors for users when choosing where to run a job. 
    }
    \label{fig:survey-selection}
\end{figure}

We first examine HPC user awareness of their resource use.  73\% of respondents (148 people) were aware of how many node-hours a job/workflow consumes, and 70\% (142) indicated that they had taken steps to reduce the number of node-hours used. This aligns with the more than 80\% of respondents (166) who were very or mildly concerned with finishing their jobs within their allocation, typically measured in node-hours. 77\% of those concerned about completing their jobs had taken steps to reduce the node-hours their jobs consume. \textbf{Overall, users are aware of their node-hour usage, concerned about completing their jobs within their allocation, and take steps to reduce their node-hours as a result.}  

In contrast, only 27\% of respondents (51) were aware of the amount of energy their workload consumes and 30\% (54) had taken steps to reduce energy use. Counterintuitively, there is not strong overlap between these groups. 39\% of people who have taken steps to reduce their energy use were not aware of how much energy their jobs consume. 
\textbf{Few users are aware of their energy consumption or have taken steps to reduce it.}

We also examine familiarity with metrics that assess efficiency. Rankings like Green500~\cite{green500} and Graph Green500~\cite{hoefler-greengraph500}, and benchmarks like the SPEC Server Efficiency Rating Tool (SERT)~\cite{lange-sert}, distill machine efficiency to a single metric to help providers or users select resources. Carbon intensity captures carbon emissions from producing electricity, which varies based on facility location~\cite{electricity-maps}. Power Usage Efficiency (PUE) captures the fraction of facility electricity that is used for computing rather than cooling or other needs~\cite{amazon-sustainability}.  Participants reported some familiarity with these metrics, with 51\% (94) and 30\% (55) of respondents knowing of Green500 and carbon intensity, respectively. However, this knowledge does not affect user behavior. As shown in \autoref{fig:survey-tools}, few respondents were aware of how the resources they use perform on these metrics. For instance, of the 94 people familiar with the Green500 list, only 36 (20\% of all respondents) knew how the machine they were using performed on that ranking. \textbf{While users are familiar with metrics designed to improve sustainability, most do not know how they apply to their own machines.}

Given these findings, it is not surprising that energy efficiency is among the least important metrics respondents used when selecting a machine. 
In most cases users have many options when choosing which machine to use: more than 70\% of users have access to four or more machines.
\autoref{fig:survey-selection} shows participant responses when asked how important various parameters were in selecting which machine to use.
While 46\% of respondents (83) said machine performance was very important, only 12\% (25) said energy efficiency was very important. This is a challenge for providers looking to invest in more efficient options for users: without a shift in incentives, users will prioritize performance over efficiency. \textbf{In summary, users do not select machines based on energy efficiency, instead they prioritize hardware availability, queue times, performance and funding.}


\section{Impact-Based Accounting}
Our survey revealed that most users of HPC resources do not consider energy (or carbon) when deciding what to run or where to run it. 
To alleviate this issue, we explore resource accounting models in which the cost of using a machine is based on the energy or carbon consumed by the computation.
As we discuss in the following, our approach combines solutions to two problems:
(1) How to incorporate both energy and usage into an accounting method; and
(2) How to account for carbon.

\subsection{Background: Fungible Allocations}
We design accounting methods for \textit{fungible} allocations: allocations that are resource independent and 
may be redeemed on multiple resources. 
Such allocations are employed, for example, by the US National Science Foundation (NSF) Advanced Cyberinfrastructure Coordination Ecosystem: Services and Support ACCESS~\cite{boerner-access},
Chameleon Cloud~\cite{keahey-chameleon}, and internally within Google~\cite{tirmaze2020borg}. Fungible allocations provide flexibility by allowing users to select between different machines, which may potentially have different efficiencies. Fungible allocations may be based only on time (e.g., Chameleon Cloud allocates users a number of node-hours that can be redeemed on any node type) or on a mix of time and performance (e.g., ACCESS grants users \textit{service units} that can be exchanged for allocations on specific machines based on machine-specific exchange rates; Google tracks \textit{Google Compute Units} which standardizes core-time to the same amount of computational power on any machine in the fleet).

\subsection{Energy-Based Accounting Model}\label{sec:energy_allocation_model}
Our \textbf{Energy-Based Accounting (EBA) }model \textit{charges users for energy used rather than time spent computing}. 

Intel and AMD CPUs support accurate CPU energy readings~\cite{kahn-rapl-2018} that can be collected by a cluster management tool~\cite{slurm-energy-acct}. 
Although not available for this work, Baseboard Management Controllers (BMCs) can on some machines be used to measure energy used by a whole node.
To account for differences in data-center design and cooling, the measured energy could be multiplied by the PUE.
Many HPC centers already implement monitoring of power consumption~\cite{georgiou-scheduler-incentive, solorzano2024toward}, so these energy could be incorporated into accounting models without monetary overhead.

A challenge with charging for energy consumed is the trade-off between \textit{potential} and \textit{actual} usage.
When accounting for time, i.e., node-hours, providers charge for the entire allocated resource, regardless of how it is used, because the resource cannot also be allocated to another user.
That is, they charge based on potential rather than actual usage. 
An accounting model that charges based on energy consumed rather than time spent results in the end price depending on user activity.
This is a double-edged sword: users who write more efficient software are rewarded with reduced costs, but so are users who do not fully use allocated hardware, even though the provider cannot charge others for the resource.

To address this issue, we incorporate the potential use of a node into our calculation. 
We use a processor's Thermal Design Power (TDP)---the maximum sustained power that it can dissipate---as a surrogate for a node's full utilization.
We charge users for the average of the actual energy used and the energy that would have been consumed had they used the resource fully. For a job $j$ on resource $R$ that uses energy $e_j$ and takes duration $d_j$, the cost based on the \textbf{energy charge} is:
\begin{equation}
\label{eqn:energy-credit}
    \hat{e}_j =  (e_j + \cdot d_j \cdot \textnormal{TDP}_R)/2.
\end{equation}
As a refinement, the second term could be weighted by a parameter $\beta<1$ for scenarios where the TDP of a device is much greater than the typical power use, skewing the cost. However, we do not employ that refinement here. 

Factoring potential use into EBA, ensures that energy efficiency is balanced with job duration. 
The average between energy used and TDP is both simple and transparent.



\subsection{Carbon-Based Accounting Model}\label{sec:carbon}

Energy consumption is not the only factor that contributes to the sustainability of a computing facility~\cite{baolin-sustainable-hpc}. Estimating the carbon footprint of a computation gives us a mechanism to account for more holistic impacts. To that end, our \textbf{carbon-based accounting (CBA)} model charges users based on the gCO$_2$e, i.e., grams of carbon dioxide equivalent, used.
This footprint encompasses the \textit{operational carbon} of the electricity used to run a job and a portion of the machine's \textit{embodied carbon} that we attribute to a job. 
%

We use the carbon intensity, $I_f(t)$, of the electricity grid at facility $f$ at time $t$ to estimate the operational carbon of a job.   
Carbon intensity captures the carbon emissions produced in generating electricity, in gCO$_2$e per kWh. This measure depends on the generation source, which varies by location and time; estimates can be obtained from grid operators or public APIs~\cite{electricity-maps}.

Embodied carbon provides a principled method of weighting potential use of a resource against actual use.
To attribute embodied carbon to a job, we differ from the standard practice of allocating the embodied carbon of a device linearly based on time~\cite{sci}. Instead we treat the embodied carbon like a capital expense invested in the machine that depreciates over time. Our rationale for this approach is that the embodied carbon allocated to a job should be proportional to the utility derived from using the machine. 
Users demanding the latest technologies drive providers to acquire new machines. Alternatively, users who leverage older technology allow hardware providers to extend refresh cycles~\cite{lyu2023myths}. Thus, our model charges a higher rate to users who use a machine earlier in its lifespan. 

Specifically, we employ a form of accelerated depreciation called double declining balance~\cite{irs-depreciation}. 
In line with typical refresh periods~\cite{lyu2023myths}, we assume an HPC machine has a depreciation period of five years, which corresponds to a 40\% annual depreciation rate.
For a machine with $C_f$ total embodied carbon,
the unaccounted-for carbon $y$ years after the machine was installed is:
$$R_f(y) = C_f \cdot (1-0.4)^y;$$ 
the embodied carbon allocated to year $t$ is: 
$$D_f(y) = 0.4 \cdot R_f(y);$$ 
and the carbon-rate per hour of resource utilization is $$D_f(y)/(24 * 365).$$
Thus the total \textbf{carbon charge} for a job $j$ run at facility $f$ with carbon intensity at time $t$ of $I_f(t)$ that uses $e_j$ kWh is:
\begin{equation}
\label{eqn:carbon-credit}
    c_j = e_j \cdot I_f(t)  + d_j \cdot D_f(y)/(24 * 365).
\end{equation}


\section{\Sysname{} System Implementation}

We implemented \sysname{}, a HPC-FaaS platform that provides fungible allocations across machines with EBA and CBA. 

\subsection{\Sysname{} Design}
\begin{figure}[t]
    \centering
    \includegraphics[width=\linewidth]{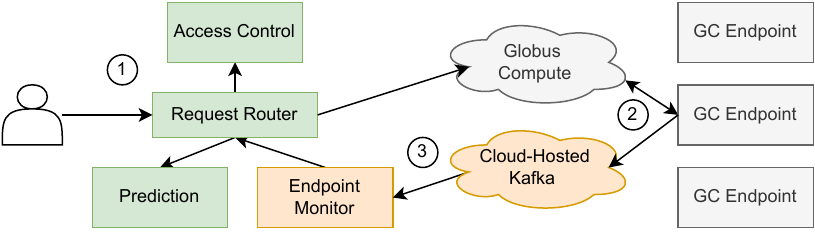}
    \Description[A flow chart of the \sysname{} prototype.]{The flow chart is divided into 3 big areas: the front end with the access control, the backend of Globus Compute, and the monitoring framework.}
    \caption{Architecture of the \sysname{} prototype. The three principal system components, shaded green, grey, and orange, are described in the text.}
    \label{fig:prototype-design}
\end{figure}

We implement \sysname{} with a FaaS interface that provides a flexible and adaptive runtime, allowing us to efficiently realize impact-based allocations. Recent work shows that FaaS can improve resource utilization and accelerate applications on HPC systems~\cite{chard-funcx,copik-rfaas,kamatar2024greenfaas}.
FaaS simplifies migration of applications between different systems which we leverage to show the differences between impact-based accounting and other accounting mechanisms.
Our system comprises three major components (see \autoref{fig:prototype-design}).

\circled{1} The frontend supports user interactions,
with accounting and admission control.
Users can employ a Web 
or Python interface to access a prediction service that provides estimates of the energy consumption of their jobs.
Jobs are submitted via a FaaS interface.

\circled{2} 
\Sysname{} uses Globus Compute as the FaaS platform 
to run Python functions on HPC systems~\cite{chard-funcx}. \sysname{} calculates expected costs before forwarding requests to Globus Compute. 
Registering a machine with \sysname{} requires deploying a Globus Compute Endpoint (GCE) equipped with a monitor that polls data from the RAPL interface, reads hardware counters, and communicates those data back to \sysname{}. This integration with Globus Compute allows \sysname{} to be deployed within (or beside) existing cluster management systems. 

\circled{3} Energy and performance counter data are transferred via Kafka to \sysname, where they are 
consumed by the \sysname{} endpoint monitor, a streaming consumer based on the Faust library~\cite{faust}. This monitor disaggregates per-node power measurements from the RAPL subsystem~\cite{david2010rapl, kahn-rapl-2018} into user jobs, which are provisioned by core. To this end, we collect per-process hardware performance counters and periodically fit a power model between performance counters and measured energy~\cite{func-energy-andre, smart-watts}.  Per-process estimates are aggregated to obtain the energy used by a task. 
For GPUs, we assume that an entire GPU is allocated to each job.


\subsection{Comparison of Accounting Methods}

Using \sysname{}, we evaluate the price of executing functions on CPUs and GPUs with five accounting methods:
\begin{itemize}
    \item \textbf{Runtime}: Price is determined only by the core-time used, not accounting for heterogeneity. 
    This is similar to the model used by Chameleon Cloud~\cite{keahey-chameleon}.
    \item \textbf{Energy}: Price is determined only by the energy used, without accounting for device capacity.
    \item \textbf{Peak}: Price is determined by core-time used, multiplied by machine peak performance. 
    This metric accounts for heterogeneous devices by charging more for higher performance systems,
    similar to ACCESS~\cite{boerner-access}.
    \item \textbf{EBA}: The proposed Energy-Based Accounting.
    \item \textbf{CBA}: The proposed Carbon-Based Accounting.
\end{itemize}
\Sysname{} allows us to calculate the cost of an invocation using any of the accounting methods.

\subsubsection{\Sysname{} on CPUs}

\begin{figure}[t]
    \centering
    \includegraphics[width=\linewidth]{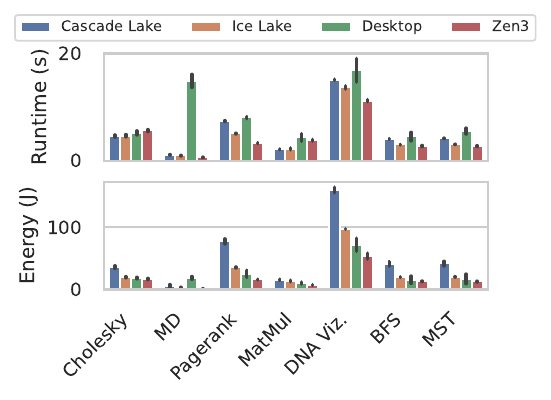}
    \Description{A bar chart with 7 different functions showing a variety of trade-offs between time and energy.}
    \caption{Runtime and energy consumption of seven applications running on four different nodes reveal different performance/energy tradeoffs across machines.}
    \label{fig:apps}
\end{figure}

We start by executing five applications from the SeBS~\cite{copik2021sebs} benchmark and two scientific applications~\cite{ward2021design, pauloski2024taps} on CPU nodes with \sysname{}. 
We select nodes of different generations and with varied performance characteristics: A Desktop with an i7-10700 CPU, a \textit{Cascade Lake} node with 2 Intel Xeon 6248R, and an \textit{Ice Lake} node with 2 Intel Platinum 8380 CPUs, and a \textit{Zen3} node with 2 AMD EPYC 7763 processors.
\figurename~\ref{fig:apps} shows the runtime and energy consumed for each application and highlights the different energy/performance tradeoffs that exist between applications and nodes.

\autoref{tab:credit-comparison} shows the normalized costs of the Cholesky decomposition application for each accounting method. We chose this application as it most clearly demonstrates unexpected tradeoffs and differences between accounting methods. 
The table shows the undesirable properties of the \textit{Runtime} and \textit{Peak} accounting methods.
As the Cholesky application runs fastest on the Cascade Lake and Ice Lake systems, charging based solely on \textit{Runtime} assigns the lowest cost on those machines. 
However, in this case, those machines also use the most energy. 
Weighting by peak performance, as in \textit{Peak} does not rectify this.
In this case, Zen3 and Desktop have the highest peak performance per thread~\cite{passmark} but are the slowest systems for this task. As a result, the Cascade Lake machine is the cheapest under the Peak accounting method, even though it uses more than 2$\times$ as much energy as the Zen3 machine. \textbf{The \textit{Runtime} and \textit{Peak} accounting methods both indirectly incentivize higher energy use.}

\begin{table}
    \setlength{\tabcolsep}{4pt}
    \centering
    \caption{Runtime, energy consumption, and costs on different CPU nodes running a Cholesky Decomposition.}
    \begin{tabular}{rcrrcrrr}
        \toprule
        Machine &\phantom{} & \multicolumn{2}{c}{Metrics} & \phantom{} & \multicolumn{3}{c}{Normalized Costs} \\
        \cmidrule{3-4} \cmidrule{6-8}
        {} && \parbox[r]{1cm}{\raggedleft Runtime (s)} & \parbox[r]{1cm}{\raggedleft Energy (J)}&& EBA & CBA & Peak \\
         \midrule
         Desktop && 5.20 & 18.3 && \textbf{1.0} & \textbf{1.0} & 1.43\\
         Cascade Lake && 4.68 & 35.8 && 1.90 & 1.20 & \textbf{1.0} \\
         Ice Lake && \textbf{4.60} & 19.8 && 1.10 & 1.10 & 1.06\\
         Zen3 && 5.65 & \textbf{16.8} && 1.05 & 1.15 & 1.36\\
         \bottomrule
    \end{tabular}
    \label{tab:credit-comparison}
\end{table}

In contrast, the \textit{EBA} price mirrors task energy usage. 
Desktop, which has the 2nd lowest energy use, has the lowest \textit{EBA} cost.
Zen3 uses the least energy but has a slightly higher \textit{EBA} cost than Desktop due to pricing for utilization in \autoref{eqn:energy-credit}: Zen3 has a higher TDP than Desktop, so it costs more to use per time. 
\textit{CBA} shares some properties of EBA, with Desktop having the lowest cost and Cascade Lake the highest. However, the newer Zen3 is charged a higher price for embodied carbon. 
\textbf{Overall, we see that running on a more efficient machine costs less under \textit{EBA} and \textit{CBA}.}


\subsubsection{\Sysname{} on GPUs}

The accounting methods and trade-offs that we have demonstrated
extend to GPUs.
We run a tiled Cholesky decomposition on a 42~GB single precision matrix, using the StarPU runtime system to orchestrate the application across different Nvidia GPUs~\cite{starpu} (see \autoref{tab:gpu-specs} for specifications).
\autoref{tab:credit-comparison-gpu} shows the energy consumed by different generations and numbers of GPUs.
We observe that:
(1) Energy consumption decreases as we scale up to four GPUs, but then stabilizes from four to eight GPUs as the application problem size is too small to saturate all GPUs. Furthermore, we see that both runtime and energy remain similar, implying good scaling performance with respect to energy. 
(2) Recent GPUs consume more energy for modest performance gains (specifically between the V100 and A100).
For the Cholesky application, the critical path and data transfers limit the benefits of the additional cores available on newer GPUs.
Indeed, the newest GPU (A100) solves the problem 6\% faster than the previous generation (V100), but consumes 60\% more energy.

\autoref{tab:credit-comparison-gpu} also illustrates how \textit{EBA} and \textit{CBA} manage this trade-off compared to other accounting methods. Eight A100 devices provide the best performance, but consume twice the energy of the P100s. On the other hand, \textit{EBA} and \textit{CBA} both prioritize using two P100 GPUs, which provide the lowest energy consumption and have the lowest embodied carbon rate compared to newer GPUs. Meanwhile, a \textit{Peak} charges the least for using one P100 GPU even as it uses 40\% more energy and 66\% more time than two P100 GPUs. 
\textbf{
EBA and CBA intentionally balance the trade-offs between the higher performance of modern GPUs and the lower energy/carbon use of older GPUs.}


\begin{table}[t]
    \centering
    \caption{Specifications of GPU nodes and carbon rate for different number of GPUs. GFlop/s is manufacturer reported. The average carbon intensity of all nodes was 53 gCO$_2$e/kWh. Embodied carbon was calculated using SCARIF~\cite{ji24scarif}.}
    \begin{tabular}{rcrrrrrrr}
        \toprule
        GPU & \phantom{} & Year & GFlop/s &  TDP & \multicolumn{4}{c}{Carbon Rate} \\
        \cmidrule{6-9}
        \# GPUs & & & & & 1 & 2 & 4 & 8\\
        \midrule
        P100 & & 2018 & 6700 & 250 & 8.5 & 9.1 & & \\
        V100 & & 2019 & 14000 & 250 & 19 & 20 & 23 & 28\\
        A100 & & 2021 & 18000 & 400 & 87 & 93 & 106 & 131\\
        \bottomrule
    \end{tabular}
    \label{tab:gpu-specs}
\end{table}

\begin{table}[t]
    \centering
    \caption{Comparison of runtime, energy use, and costs under different accounting schemes for Cholesky Decomposition with different types and number of Nvidia GPUs.}
    \begin{tabular}{rrcrrcrrr}
        \toprule
        & \# &\phantom{} & \multicolumn{2}{c}{Metrics} & \phantom{} & \multicolumn{3}{c}{Normalized Costs} \\
        \cmidrule{4-5} \cmidrule{7-9}
        {} & & & \parbox[r]{1cm}{\raggedleft Runtime (s)} & \parbox[r]{1cm}{\raggedleft Energy (kJ)} && EBA & CBA & Perf. \\
         \midrule
         P100 & 1 & & 2321 & 889 && 1.20 & 1.40 & \textbf{1.0}\\
         & 2 && 1396 & \textbf{635} && \textbf{1.0} & \textbf{1.0} & 1.20\\         
         V100 & 1 && 1494 & 1316 && 1.23 & 2.07 & 1.34\\
         & 2 && 1190 & 1194 && 1.26 & 1.88 & 2.14\\ 
         & 4 && 917 & 916 && 1.25 & 1.44 & 3.30\\ 
         & 8 && 926 & 944 && 1.85 & 1.49 & 6.67\\ 
         A100 & 1 && 1405 & 2100 && 1.83 & 3.35 & 1.62\\
         & 2 && 926 & 1427 && 1.46 & 2.28 & 2.14\\ 
         & 4 && 841 & 1320 && 1.76 & 2.11 & 3.89\\ 
         & 8 && \textbf{838} & 1325 && 2.59 & 2.13 & 7.76\\ 
         \bottomrule
    \end{tabular}
    \label{tab:credit-comparison-gpu}
\end{table}

\subsection{Comparison of Embodied Carbon}
\label{sec:embodied-comp}

In~\autoref{tab:embodied_carbon_comp}, we compare different methods for accounting for carbon using the same Cholesky decomposition application as above. 
Specifically, we compare the operational carbon to two different methods of charging for embodied carbon: a linear method of depreciation where the embodied carbon rate is constant over the lifetime of the resource~\cite{sci}, and our accelerated depreciation method.

Bashir et al.~\cite{bashir2024sunkcarbon} claim that optimizing for embodied carbon has the potential to increase overall carbon emissions.
They argue that the embodied carbon cannot be reduced, so the most sustainable policy is to use the machine with the lowest operational emissions. 
This possibility is reflected in the table: the Cascade Lake node uses 2.8~mg, the most operational carbon of any node, but is attributed the least embodied carbon, 1.0~mg based on linear depreciation and 0.3~mg based on accelerated depreciation.
Including embodied carbon makes the price of the least efficient machine closer to the other machines.
However, newer hardware is acquired at least partly in response to user behavior: users that demand and use (only) the latest hardware drive providers to retire older machines~\cite{lyu2023myths}.
Thus, we argue that this observation motivates 
the use of accelerated depreciation instead of linear depreciation to allocate more embodied carbon earlier in a resource's lifespan and extend the productive lifespan of these machines.
The accelerated depreciation attributes 0.7~mg less embodied carbon than linear depreciation on the older Cascade Lake machine, but 0.3~mg more embodied carbon on the newer Zen3 machine.
As longer lifespans will lengthen refresh cycles and lessen the number of devices manufactured, accounting for embodied carbon will reduce emissions over the long run.

\begin{table}[t]
    \centering
    \caption{Comparison of operational carbon vs. different methods of attributing embodied carbon: linear depreciation, and accelerated depreciation (ours). Compared to linear depreciation, accelerated depreciation charges less for older machines.}
    \begin{tabular}{rrcrrr}
        \toprule
        Machine & Age & \phantom{} & \parbox[r]{1.6cm}{\raggedleft Operational (mgCO$_2$e)} & \multicolumn{2}{c}{\parbox[r]{1.6cm}{\raggedleft Embodied (mgCO$_2$e)}}\\
        \cmidrule{5-6}
        & &&  & Linear & Accel. \\
        \midrule
        Desktop & 3 && 2.1 & 1.5 & 0.6 \\
        Cascade Lake & 4  && 2.8 & 1.0 & 0.3 \\
        Ice Lake & 2 && 0.9 & 1.4 & 1.0 \\
        Zen3 & 1 && 1.2 & 1.3 & 1.6 \\
        \bottomrule
    \end{tabular}
    
    \label{tab:embodied_carbon_comp}
\end{table}

\section{Simulation Studies}\label{sec:simulation}

The implementation of \textit{green-ACCESS} demonstrates that trade-offs between energy consumption and performance can lead to different machines being cheaper under \textbf{EBA} and \textbf{CBA}. To analyze EBA and CBA at scale, we conduct simulations of real workloads and machines. 
We modify an existing batch simulator to use the proposed accounting methods on multiple machines~\cite{gonthier-simulator}. Our goals are to (1) understand how different choices affect the cost of running a workload under EBA and CBA, (2) examine how different accounting models translate to resources used, and (3) explore how low-carbon scenarios may affect the accounting models.

\subsection{Machines}
\begin{table*}[t]
    \centering
    \caption{Machines used for simulation. TDP=Thermal Design Power, in Watts. Idle power is for all sockets on the node.}
    \label{tab:systems}
    \begin{tabular}{rrrrrrrrrr}
        \toprule
         Machine & \parbox[r]{1.8cm}{\raggedleft Year Deployed} & CPU Model &
         \parbox[r]{1.2cm}{\raggedleft \# of Cores}
          & \parbox[r]{1.2cm}{\raggedleft CPU TDP (W)} & \parbox[r]{1.3cm}{\raggedleft Idle Power (W)} & \parbox[r]{1.7cm}{\raggedleft Carbon Rate (gCO$_2$e/h)} & \parbox[r]{1.9cm}{\raggedleft Avg. Carbon Intensity (gCO$_2$e/kWh)}\\
        \midrule
        TAMU FASTER & 2023 & 2 $\times$ Intel Xeon 8352Y & 64 & 205 & 205 & 105.2 & 389 \\
        Desktop & 2022 & Intel Core i7-10700 & 16 & 65 & 6.51 & 12.2 & 454 \\
        Institutional Cluster & 2021 & 2 $\times$ Intel Xeon 6248R & 48 & 205 & 136 & 16.7 & 454 \\
         ALCF Theta & 2017 & Intel KNL 7320  &  64 & 215 & 110 & 2.0 & 502 \\
         \bottomrule
    \end{tabular}
\end{table*}

We characterize in \autoref{tab:systems} the machines used in our simulation. 
For each, we indicate: Cores per Node; CPU Thermal Design Power (the maximum heat in watts that a CPU can dissipate); Idle Power (power consumed by the CPUs when running only the monitoring code); 
Carbon Rate (see \autoref{sec:carbon}, where the embodied carbon was calculated using manufacturers datasheets where available or SCARIF~\cite{ji24scarif}).
For carbon intensity, we retrieved data at an hourly resolution assuming the simulation starts in January 2023, and report the yearly average.
Three of the machines are in HPC centers: FASTER a cluster at Texas A\&M University available through NSF ACCESS~\cite{boerner-access}, Midway an Institutional Cluster (IC) at the University of Chicago, and Theta a machine at Argonne Leadership Computing Facility.
The fourth is a personal computer referred to here as Desktop.

\subsection{Workload}
We use a published dataset of per-job energy use from two HPC clusters~\cite{9139801}.
In order to infer the energy characteristics, we assume jobs run by the same user with the same requested resources are repetitions of the same app and have the same cross-platform characteristics.
Discarding jobs that lack an associated energy value
reduces the dataset from $\sim$84k to \num{71190} jobs. 
We repeat each execution twice to generate a workload of \num{142380} jobs.
Any job can run anywhere, except that 17\% of jobs require more cores than are available on the one-node Desktop.

The dataset that we use to construct this workload provides, for each job, energy consumption only on one machine.
We must extrapolate these data to our machines.
To begin, we assume that the reported runtime and energy consumption correspond to running on IC, the system most similar to the source dataset.
We then adapt a method to predict energy use and runtime had jobs been run on the other machines~\cite{pham-predicting-cloud}.
First, we generate realistic values for hardware performance counters (i.e., LLC Misses/sec., Instructions/sec) for each job using a Gaussian Mixture Model trained on data collected on IC. 
We then use a KNN~\cite{pham-predicting-cloud} trained on a set of benchmark applications~\cite{copik2021sebs} to estimate runtime and power consumption on the other machines.

\subsection{Policies}

To simulate user choice, we define eight machine selection \textit{policies} that select which machine to submit a job to based on the system state (queue times) and job profile (energy use, runtime, or cost).
Each simulated user is limited to one running job on a cluster at a time.
The eight policies are:
\begin{itemize}
    \item \textbf{Greedy}: Select \machine that minimizes cost, according to the accounting method (EBA or CBA) used.
    \item \textbf{Energy}: Select \machine that minimizes energy used.
    \item \textbf{Mixed}: Balance runtime and cost. Select \machine with the least allocation cost (in terms of EBA or CBA) \textit{unless} another \machine can complete the \job in half the time, in which case select that machine.
    \item \textbf{EFT} (earliest finish time): Select \machine that minimizes completion time, i.e., queue time + runtime.
    \item \textbf{Runtime}: Select \machine with shortest runtime.
    \item \textbf{Theta}, \textbf{IC}, and \textbf{FASTER}: Always select that \machinenospace.
\end{itemize}

\subsection{Results with Energy-Based Accounting}

\begin{figure*}\centering
\begin{subfigure}[t]{.3\linewidth}
    \includegraphics[width=\linewidth]{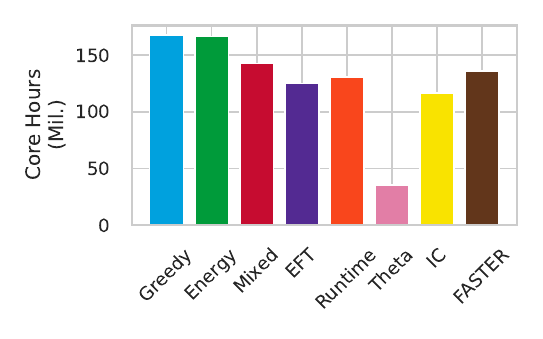}
    \Description[A bar graph showing the number of core-hours completed by each user policy.]{The bar graphs show that the Greedy and Energy policies complete more than 150 million core-hours of work. In this context core-hours is a metric of the amount of work done.}
    \caption{Work completed with fixed allocation.
    }
    \label{sim:core_hours_credit_e1}
\end{subfigure}
\hfill
\begin{subfigure}[t]{.32\linewidth}
    \includegraphics[width=\linewidth]{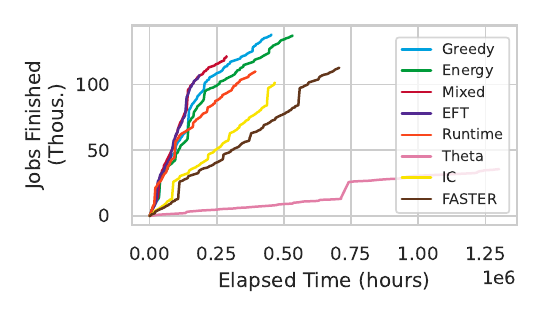}
    \Description[A line graph showing the number of jobs vs. time elapsed before allocation expired.]{There are lines for each policy. The Mixed policy and the EFT policy are the steepest policies indicating speed, but EFT completes fewer jobs.}
    \caption{Number of jobs completed over time. 
    }
    \label{sim:nb_jobs_completed_over_time_e1}
\end{subfigure}
\hfill
\begin{subfigure}[t]{.36\linewidth}
    \includegraphics[width=\linewidth]{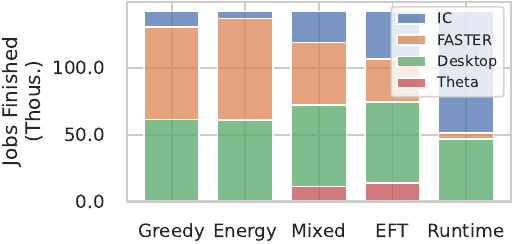}
    \Description[A stacked bar graph showing the distribution of jobs to machines by policy.]{The stacked bars show the Greedy policy and the energy policy evenly distributed tasks to FASTER and Desktop and used IC sparingly. Meanwhile, Runtime used IC primarily.}
    \caption{Distribution of \jobs over \machines by policy. 
    }
    \label{sim:stacked_e1}
\end{subfigure}
\caption{Results from simulating EBA. The \textit{Energy} and \textit{Greedy} policies complete the most work by prioritizing FASTER over IC for efficiency. The \textit{Mixed} policy trades some efficiency for faster completion time.}
\label{sim:eba}
\end{figure*}

\autoref{sim:core_hours_credit_e1} depicts the amount of work (in core-hours) completed by a user when applying each policy with a fixed allocation.
Specifically, we consider work to be the average number of core hours required to run a job across all machines.
This places higher weight on larger and longer jobs, without favoring any one machine.

In general, the single-machine policies (\textit{Theta}, \textit{IC}, and \textit{FASTER}) and policies that do not consider energy (\textit{EFT} and \textit{Runtime})
perform less work for the same cost.
A user using \textit{Theta} is severely disadvantaged by EBA because they submit all jobs to a machine that is inefficient for most tasks.
Using \textit{Greedy}, a user is able to complete the most work, because, by definition, they use the cheapest machine for every task. 
This results in completing 28\% more work than when using the performance focused \textit{EFT} policy.
A user employing \textit{Energy} can complete 99\% of the work done using  \textit{Greedy} because the most efficient machine is often the cheapest machine.
\textbf{Under EBA, by optimizing for energy, a user is indirectly optimizing for cost; in other words, the cost incentivizes users to be more energy efficient.} 

We now consider only the multi-machine policies and examine how each policy affects total energy consumption.
\autoref{tab:energy_carbon_used_carbon_credit} shows the energy consumed using each policy over the workload.
Logically, we see that using \textit{Energy} consumes the least energy.
A user optimizing cost instead of the energy by using \textit{Greedy} resulted in only 2\% more energy use.
In contrast, applying \textit{EFT} or \textit{Runtime} results in 51\% or 56\% more energy used, respectively.
Much like we saw in real hardware, performance and efficiency are not always aligned. 
\textbf{A user prioritizing the speed in terms of either makespan or core-hours may make inefficient decisions.}

\begin{table}[t]
    \centering
    \caption{Energy and carbon used when deploying each policy while computing the workload. We show how the \textit{Greedy} and \textit{Mixed} policy behaved both under \textit{EBA} and \textit{CBA}.}
    \label{tab:energy_carbon_used_carbon_credit}
    \begin{tabular}{rrrr}
         \toprule
         Policy &  Energy  & \multicolumn{2}{c}{Carbon (KgCO$_2$e)}\\
         &  (MWh) & Operational & Attributed\\
         \midrule
         \textit{Greedy - EBA} & 328 & 88 & 322\\
         \textit{Greedy - CBA} & 491 & 167 & \textbf{228}\\
         \textit{Mixed - EBA} & 407 & 132 & 319\\
         \textit{Mixed - CBA} & 494 & 172 & 275\\
         \textit{Energy} & \textbf{321} & \textbf{83} & 345\\
         \textit{EFT} & 486 & 169 & 315\\
         \textit{Runtime} & 501 & 170 & 237\\
         \bottomrule
    \end{tabular}
\end{table}

Next, we examine how the different policies relate to completion time. Using a single machine is detrimental in terms of completion time because of long queue times. This is visible in \autoref{sim:nb_jobs_completed_over_time_e1},
which shows the jobs completed by different policies over time.
We also see that using \textit{Mixed} a user is able to compute \num{100000} jobs as fast as using \textit{EFT}.
\textbf{By balancing energy efficiency and performance, users can reduce cost under EBA without impacting completion time.}

Finally, we investigate how EBA might affect the usage of each machine. \autoref{sim:stacked_e1} shows how jobs are distributed with each policy.
\textit{Greedy} and \textit{Energy} policies distribute jobs similarly to FASTER, Desktop, and IC.
\textit{Greedy} and \textit{Energy} allocate no tasks to Theta because it is neither the cheapest nor the most energy efficient.
However,  \textit{Mixed} distributes tasks to all four machines to avoid long queues and reduce the completion time.
\textbf{Overall, EBA may lower utilization of inefficient machines to reduce energy consumption, but users may choose to use less efficient machines at a higher cost to reduce completion time.}

\subsection{Results with Carbon-Based Accounting}

\begin{figure}
\centering
    \includegraphics[width=.8\linewidth]{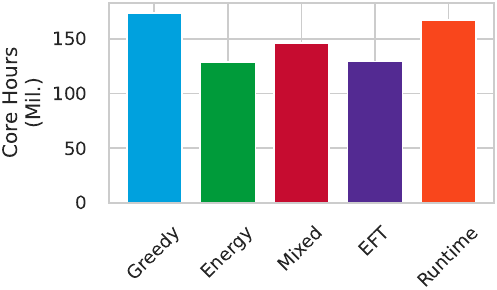}
    \Description[A bar graph showing the work completed under a CBA allocation.]{The bar graph shows that the Greedy policy completed the most work (more than 150 million core-hours); runtime and mixed had the next most.}
    \caption{Results from a simulating of CBA, showing the amount of work a user completed for a fixed cost by employing different machine selection policies . 
    }
    \label{sim:mean_completion_carbon}
\end{figure}

\begin{figure*}
\centering
\begin{subfigure}[t]{.31\linewidth}
    \includegraphics[width=\linewidth]{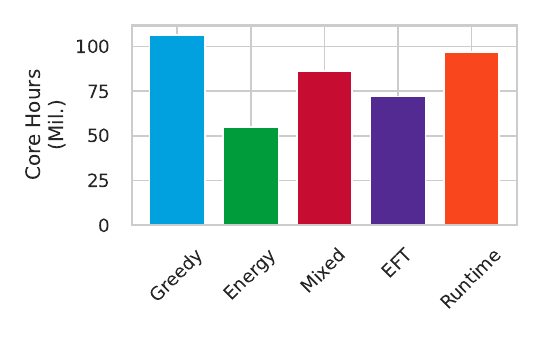}
    \Description[A bar graph showing the amount of work completed with a CBA allocation in a reduced carbon intensity scenario.]{The bars show that Greedy is still able to complete more than 150 Million core-hours while the next closest is Runtime with approximately 125 million core-hours.}
    \caption{Work completed with a fixed CBA allocation and with reduced carbon intensity.}
    \label{sim:mean_completion_carbon_e2}
\end{subfigure}
\hfill
\begin{subfigure}[t]{.33\linewidth}
    \includegraphics[width=\linewidth,width=\linewidth]{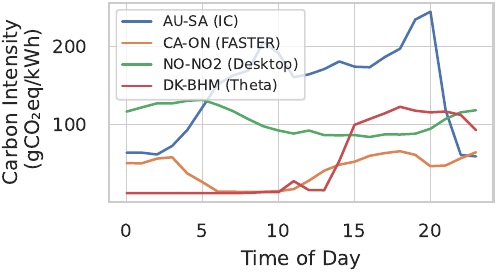}
    \caption{Variations in carbon intensity during a single day of the simulation.}
    \label{sim:varying_carbon_intensity}
\end{subfigure}
\hfill
\begin{subfigure}[t]{.32\linewidth}
    \includegraphics[width=\linewidth]{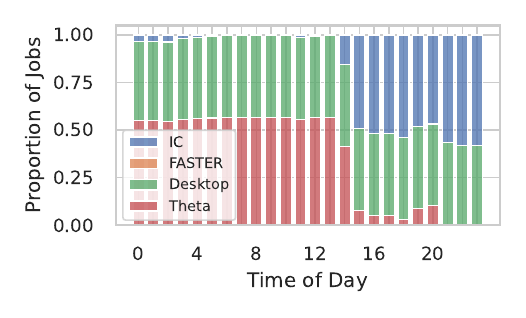}
    \Description[A stacked bar graph showing the distribution of tasks in a reduced carbon scenario.]{The stacked bars show the greedy policy prefers Desktop, IC, and Theta but does not use FASTER.}
    \caption{Distribution of of the lowest cost endpoint as the carbon intensity varies.}
    \label{sim:stackedbarplots_e2}
\end{subfigure}
\caption{Results from simulating CBA under a low-carbon scenario. As the carbon intensity varies, the cheapest machine for a job can change. Efficiency becomes less important when carbon intensity is low, so Theta is often the cheapest machine.}
\end{figure*}

Next, we analyze CBA as described in \autoref{sec:carbon}.
For \autoref{sim:mean_completion_carbon}, we allow a user employing \textit{Greedy} to run the same amount of work as in \autoref{sim:core_hours_credit_e1}.
We see that using \textit{Energy}, a user can run 22\% fewer jobs than under EBA, while using \textit{Runtime}, a user can run 23\% more. 
This is due to the high carbon rate of FASTER: while often selected by \textit{Energy} because it is the most efficient machine (see \autoref{sim:stacked_e1}), it has a much higher embodied carbon rate. 
Since \textit{Energy} favors FASTER and \textit{Runtime} favors IC, deploying \textit{Runtime} policy allows a user to complete more work in the same allocation. \textit{Greedy} adapts to the new accounting scheme and submits 50\% of its workload to IC and only 11\% 
to FASTER.

The energy and carbon consumed while using each policy are detailed in \autoref{tab:energy_carbon_used_carbon_credit}.
In the table, we decompose the operational carbon emissions (that is, the carbon intensity of the grid multiplied by the energy used), from the attributed carbon emissions, which includes a portion of the embodied carbon of the machine.
Based off the attribution of CBA, the direct carbon emissions are between 24\% and 72\% of the total carbon attributed to a job.
In particular, \textit{Energy} shows the largest share of its carbon footprint attributed to embodied carbon as a result of heavily using the latest hardware.
\textbf{Minimizing cost allows a user to compute the most work while using the least attributed carbon, incentivizing the selection of \textit{efficient} and \textit{older} machines.}

\subsection{Results with Reduced Carbon Intensity}

As countries transition to green electricity sources, grids have a higher temporal variation in carbon intensity~\cite{fan-zcc}.
We simulate a scenario involving low-carbon grids with high temporal variation. We assign each of our systems to
a grid with high
variability in carbon intensity: Southern Australia (IC); Ontario, Canada (FASTER); Bornholm, Denmark (Theta); and Southern Norway (Desktop)~\cite{sukprasert2023quantifying, electricity-maps}.
We do not adjust the embodied carbon rate in this simulation.

In \autoref{sim:mean_completion_carbon_e2}, we show the work completed by each policy under this simulation. As before, the carbon-aware \textit{Greedy} completes significantly more work than the other, non-carbon-aware policies.
The interesting difference is in how \textit{Greedy} exploits the variability in carbon.
Throughout the day, the carbon intensity varies with the electricity source in each region. The carbon intensity on an arbitrary day in the simulation is shown in \autoref{sim:varying_carbon_intensity}. 
This variation affects the distribution of the cheapest cost endpoint for each job: see \autoref{sim:stackedbarplots_e2}.
The cheapest location shifts from Theta to IC later in the day, as carbon intensity increases in Denmark and renewable generation comes online in Southern Australia. 
In the simulation (as well as above), we do not allow job migration: once a job has been started on a machine, it cannot move even as the carbon intensities change.
\textbf{CBA incentivizes users to spatially and temporally align their jobs with periods of renewable energy generation.}

\section{User Study}
In this final study we examine whether and how user behavior changes 
under a modified accounting method. 

\subsection{Design}
\begin{figure}
    \centering
    \includegraphics[width=\linewidth]{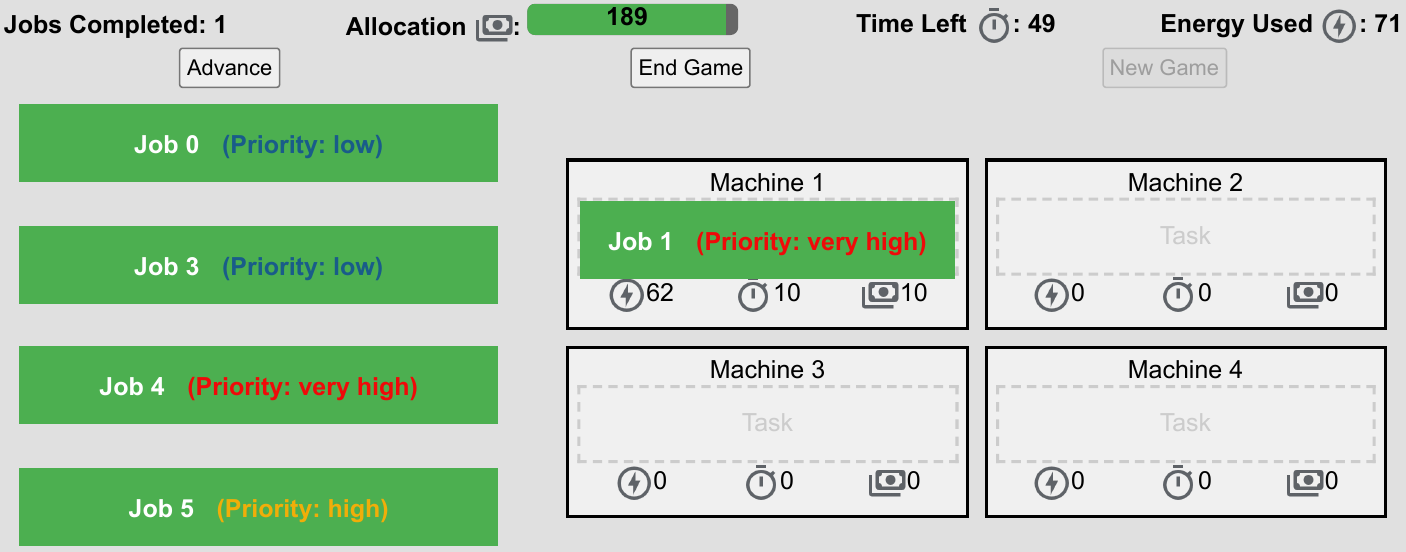}
    \Description[A screen shot of the scheduling game.]{The scheduling game shows a set of jobs on the left of the screen that can be dragged and dropped onto four different grey boxes corresponding to different systems.}
    \caption{Scheduling game. The green ``job'' boxes can be ``scheduled'' by dragging them to the light-grey ``machine'' boxes. Job time and cost, was shown by hovering over a job.}
    \label{fig:game-interface}
\end{figure}

We built a JavaScript game to emulate decisions faced by users of a \sysname{}-like platform: see \autoref{fig:game-interface}. Participants were asked to imagine that they were a computational scientist who had to finish all of their jobs within a time and allocation limit, and they had four machines to use to do so.
These instructions are a proxy for realistic HPC use cases where a researcher is granted a limited allocation but faces pressure to meet a deadline. 
In our game, a job could be ``scheduled'' by dragging and dropping it onto a specific machine. 
Every job was randomly assigned one of four priorities between ``Low'' and ``Very High.''
The priority functioned as a placebo metric. No instructions were given on how to treat priorities, so users had to weigh priority against a jobs time, energy, and cost.
The machines reflected those used in the simulation, and the resources a job used were inferred using the same mechanism as the simulation.  
More jobs ``arrived'' as jobs were scheduled, to reflect the time-dependent nature of real workloads. 
The jobs were the same for all participants.
Each participant was randomly assigned to one of three game versions:
\begin{itemize}
    \item \textbf{V1}: Job cost is proportional to runtime, and no information on job energy consumption is shown. This reflects current standard practice: see \autoref{sec:survey-results}.
    \item \textbf{V2}: Cost is the same as V1, but job energy consumption is displayed next to time and cost.
    \item \textbf{V3}: Cost is given based on the EBA formula.
\end{itemize}
Participants played the game at least twice, with the first iteration, intended for familiarization with the game mechanics. 
The version remained the same between the first and second play of the game, but was randomized after that. The game was distributed using the same channels as the survey.

\subsection{Results}
After discarding every users first time playing the game, we received 207 instances of the game played by 90 unique users. 
We discarded 15 instances in which participants finished the game in less than one minute as these also had a disproportionately high amount of allocation remaining.

\begin{figure*}
    \centering
    \hfill
    \begin{subfigure}[t]{.32\textwidth}
        \includegraphics[width=\linewidth,trim=2.5mm 2.5mm  2.5mm 2.5mm,clip]{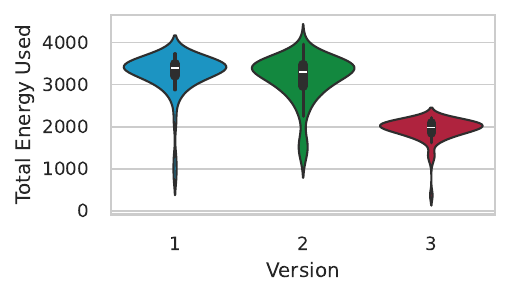}
        \caption{Energy used across different versions of the game. On average, participants used significantly less energy in V3 (with EBA).}
    \end{subfigure}
    \hfill
    \begin{subfigure}[t]{.32\textwidth}
        \includegraphics[width=0.97\linewidth,trim=2.5mm 3mm 2.5mm 2.5mm,clip]{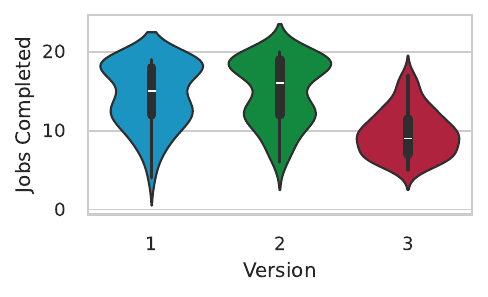}
        \caption{Jobs completed with different versions. 
        V3 participants ran fewer jobs, despite allocation intended to be equivalent to V1 and V2.
        }
    \end{subfigure}
    \hfill
    \begin{subfigure}[t]{.32\textwidth}
        \includegraphics[width=\linewidth,width=\linewidth,trim=2.5mm 2.5mm 2.5mm 2.5mm,clip]{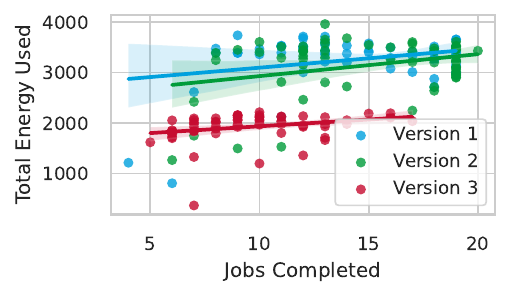}
        \caption{Energy use by the jobs completed per user across versions. For any given number of jobs completed, users in V3 used less energy.}
        \label{fig:game-energy-by-jobs}
    \end{subfigure}
    \hfill
    \Description{3 Graphs showing an analysis of the user study.}
    \caption{Comparison of the energy used during the three different versions of the game.}
    \label{fig:game-energy-runtime}
\end{figure*}

We first assess the impact of displaying energy information (V2) and using EBA (V3) on total energy consumption: see \autoref{fig:game-energy-runtime}. On average, participants using EBA (V3) consumed 1928 kWh in the simulation, while those with V1 or V2 consumed 3262 kWh and 3142 kWh, respectively. V3 is significantly lower than V1 or V2 (p=0.00), but there was no significant difference between V1 (the control) and V2 (with energy information). \textbf{Information regarding energy consumption alone created no change in how much energy a participant consumed.} 


However, in V3 (using EBA), participants also completed fewer jobs overall.  Users with the changed allocation scheme completed an average of 9.7 jobs compared to 14.5 in V1 and 14.9 in V2. Although we attempted to give an equivalent sized allocation to users in V3 as V1 and V2, the differing accounting method meant an exact conversion was impossible. To distinguish the effect of EBA from a smaller allocation, in \autoref{fig:game-energy-by-jobs} we examine the amount of energy used, stratified by the number of jobs completed. The figure shows for any number of jobs completed, \textbf{users who saw an EBA version of the game used significantly less energy}. 

\begin{figure}
    \centering
    \includegraphics[width=.9\linewidth,trim=2.5mm 2.5mm 2.5mm 2mm,clip]{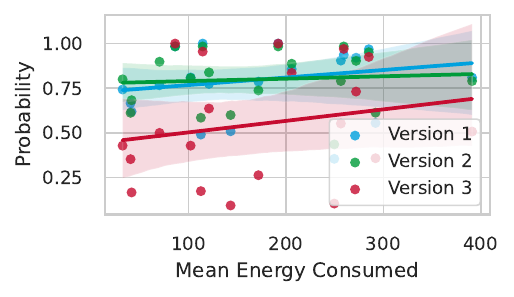}
    \Description[A plot of the probability a job was run vs. the energy consumed by version of the game.]{The plot shows near horizontal lines with wide uncertainty regions showing no correlation between energy consumed and probability.}
    \caption{Proportion of participants who ran a job vs.\ average energy consumed by job.  Energy use was not correlated with probability of running a job in any version of the game.}
    \label{fig:game-job-prob-energy}
\end{figure}

To use less energy while running the same number of jobs, a user had two options: run less energy intensive jobs, or use more efficient machines to run the same jobs. In order to distinguish these two mechanisms, we looked at the probability that a user selected to run a job. Since users may have ran out of time or allocation before seeing a job, we calculate the probability as (\# of users who completed job $i$ $/$ \# of users who saw job $i$).  If participants reduced energy by not running energy intensive jobs, we would see a (negative) correlation between the probability a job was run and its energy use. We plot this correlation in \autoref{fig:game-job-prob-energy}. While V3 participants were less likely to run a job in general (since they completed fewer jobs on average), there is no correlation between job energy use and the probability that a user ran a job. Thus, \textbf{the decision to run a job was not influenced by  energy consumed, even when cost depended on energy.}

The second mechanism for consuming less energy was to employ more efficient resources. 
We assess the efficiency of resources used by measuring the average energy used to run a job by all participants playing a specific version of the game.  
Lower average energy consumption means more participants chose more efficient machines. 
For 16 of the 20 jobs, the average energy used by participants in V3 of the game was the lowest of the three versions. This suggests that \textbf{under EBA, when participants elected to run a job, they selected a more efficient machine to do so.}

\section{Limitations}
We assumed that applications are able to be migrated between machines, and providers will cooperate to enable efficient and sustainable use of computing.  Here, we address the validity and limitations of these assumptions.

\paragraph{Reliance on Portability}
Performance portability is an open challenge and our results are further evidence of the need for portability solutions.
Many HPC users are agnostic to underlying machine details. 
For instance, users writing high-level PyTorch code can easily move between most GPU devices without modifying their code.  
In such cases, selecting between machines based on energy or carbon can already yield substantial reduction.
Ongoing research into libraries that abstract different machine-specific backends and containers that ease the deployment of applications will facilitate more portable applications.
Even in situations where only a single machine is available or suitable for an application, EBA/CBA can incentivize the adoption of energy-aware schedulers~\cite{app-aware-geopm, kamatar2024greenfaas} and other tools~\cite{chung2024perseus}. 
As we find in our survey, these tools have not been well adopted, possibly due to the lack of incentive structures.

\paragraph{Simplified Provider Behavior} 
Our study of EBA/CBA assumes that providers will cooperate to incentivize users to make sustainable choices. 
Since HPC systems are typically built/funded by governments/universities for the public interest, this cooperation is nominally aligned with the goals of the centers. 
However, we do not consider latent motivations, such as competition between facilities for future funding, that complicate the modeling provider behavior. 
Furthermore, EBA/CBA may increase the energy use or carbon footprint of a single machine in order to reduce the overall impact, which could make sites reluctant to adopt these approaches.
We hope incentive structures like EBA and CBA will stimulate the discussion on energy/carbon efficiency of scientific computing at a global level.

\section{Related Work}

\textbf{Motivating sustainable behavior.} 
Limited work studies user incentives and behavior in regards to HPC use and energy.
Georgiou et al.\ suggest incorporating energy into the cluster scheduling algorithm to prioritize energy-efficient users~\cite{georgiou-scheduler-incentive}.
Kishwar et al.\ propose a contract-based mechanism to provide HPC operators flexibility in scheduling jobs by rewarding users, but ignore the effects of renewable electricity generation and of heterogeneous hardware~\cite{kishwar2018contractbased}. 
Fugaku implemented a points based system to incentivize users to enable power-control knobs~\cite{solorzano2024toward}. Jobs that use less than the ``standard power'' on a node are rewarded with additional node-hours.
Di Pietrantonio et al.\ suggest that the Thermal Design Power (TDP) of a device can be used to calculate a static cost-ratio between GPU and CPU nodes~\cite{di2021energy}. 
Other works have proposed incorporating the price of electricity into the cost of cloud VMs~\cite{hinz2018cost,kurpicz2016evape,aldossary2018energycost,narayan2014metering}. These works emphasize deferring costs rather than incentivizing sustainability, and do not examine the same trade-offs as here. Margery et al.\ propose attributing CO\textsubscript{2} emissions to VMs~\cite{margery2017co2}.

\textbf{Sustainable data centers.}
Prior work examined adapting data-center capacity or demand to reduce power, cost, and carbon emissions~\cite{wei-renewable-datacenters, liu-green-balancing, sukprasert2023quantifying}, for example by scheduling jobs to data-centers currently powered by renewable energy.
The Zero-Carbon Cloud project examines running data-centers on ``stranded'' power---excess power generated when grid supply exceeds load~\cite{fan-zcc}. 
Such scheduling requires incentives for users to provide jobs that can be delayed or moved. 


\textbf{Green serverless.}
The problem of 
emissions has been noted by the FaaS community~\cite{patros2021toward, sharma2023challenges}.
GreenCourier implements carbon-aware scheduling of serverless functions based on cluster
location~\cite{chadha2023greencourier}, and GreenWhisk~\cite{serenari2024greenwhisk} extends
load balancing with the grid's carbon intensity and energy status of off-grid executors.
Sharma and Fuerst improve the accuracy of software power meters for serverless functions~\cite{sharma2024accountable}.
Lin and Mohammed similarly propose carbon aware pricing for serverless, but consider optimizations to function configuration rather than heterogeneity between systems~\cite{lin2024bridging}.
Roy et al. examine the carbon footprint of FaaS, highlighting the impact of keep-alive time~\cite{roy24hiddencarbon}, but keep-alive time is not applicable to HPC environments, Globus Compute, or \sysname.
EcoLife uses multi-generation hardware to reduce the carbon footprint of serverless platforms~\cite{jiang2024ecolifecarbonawareserverlessfunction}. While EcoLife focuses on scheduling and keep-alive time, we focus on user behavior.

\section{Summary}
The responsibility for sustainable computing is shared by producers and consumers.
Yet we find that more than 70\% of HPC users remain unaware of the energy consumed by their computations. 
With the goal of realizing shared responsibility, we introduced Energy- and Carbon-Based Accounting (EBA and CBA), two mechanisms for charging for computing based on environmental impact.
We built a prototype FaaS-HPC platform to implement EBA and CBA and demonstrated that both incentivize more sustainable decisions on real hardware.
We then used simulations to show that impact-based charging allows an energy-conscious user to process 28\% more  workload than a performance-focused user with the same allocation.
Finally, we demonstrated that users in a simulated environment use less energy under EBA than traditional time-based accounting.
\textbf{By linking charging to energy use and carbon emissions, EBA and CBA incentivize users to prioritize sustainable computing.
Thus our work identifies concrete steps that the HPC community can take to reduce its environmental impact.}

\begin{acks}
Results presented in this paper were obtained using the Chameleon testbed supported by the National Science Foundation. 
This work used FASTER at Texas A\&M University through allocation CIS230030 from the Advanced Cyberinfrastructure Coordination Ecosystem: Services \& Support (ACCESS) program, which is supported by U.S. National Science Foundation grants \#2138259, \#2138286, \#2138307, \#2137603, and \#2138296.
The GPU experiments presented in this paper were carried out using the Grid'5000 testbed, supported by a scientific interest group hosted by Inria and including CNRS, RENATER and several Universities as well as other organizations (see \url{https://www.grid5000.fr}).
\end{acks}

\clearpage

\balance
\bibliographystyle{ACM-Reference-Format}
\bibliography{references.bib}

@Article{starpu,
  author 	= {C{\'e}dric Augonnet and Samuel Thibault and Raymond Namyst and Pierre-Andr{\'e} Wacrenier},
  title		= {{StarPU: A Unified Platform for Task Scheduling on Heterogeneous Multicore Architectures}},
  journal	= {Concurrency and Computation: Practice and Experience, Special Issue: Euro-Par 2009},
  volume	= 23,
  issue		= 2,
  year		= 2011,
  publisher	= {John Wiley & Sons, Ltd.},
  doi		= {10.1002/cpe.1631}
}

@INPROCEEDINGS{9139801,
  author={Patel, Tirthak and Wagenhäuser, Adam and Eibel, Christopher and Hönig, Timo and Zeiser, Thomas and Tiwari, Devesh},
  booktitle={IEEE International Parallel and Distributed Processing Symposium}, 
  title={What does Power Consumption Behavior of {HPC} Jobs Reveal?: Demystifying, Quantifying, and Predicting Power Consumption Characteristics}, 
  year={2020},
  volume={},
  number={},
  pages={799-809},
  doi={10.1109/IPDPS47924.2020.00087},
  publisher={IEEE},
  address={New York, NY, USA},
}

@INPROCEEDINGS{app-aware-geopm,
  author={Wilson, Daniel C. and Jana, Siddhartha and Marathe, Aniruddha and Brink, Stephanie and Cantalupo, Christopher M. and Guttman, Diana R. and Geltz, Brad and Lawson, Lowren H. and Al-rawi, Asma H. and Mohammad, Ali and Keceli, Fuat and Ardanaz, Federico and Eastep, Jonathan M. and Coskun, Ayse K.},
  booktitle={IEEE International Parallel and Distributed Processing Symposium}, 
  title={Introducing application awareness into a unified power management stack}, 
  year={2021},
  volume={},
  number={},
  pages={320-329},
  doi={10.1109/IPDPS49936.2021.00040},
  publisher={IEEE},
  address={Virtual}
}

@ARTICLE{wei-renewable-datacenters,
  author={Deng, Wei and Liu, Fangming and Jin, Hai and Li, Bo and Li, Dan},
  journal={IEEE Network}, 
  title={Harnessing renewable energy in cloud datacenters: Opportunities and challenges}, 
  year={2014},
  volume={28},
  number={1},
  pages={48-55},
  doi={10.1109/MNET.2014.6724106}
}

@INPROCEEDINGS{fan-zcc,
  author={Yang, Fan and Chien, Andrew A.},
  booktitle={IEEE International Parallel and Distributed Processing Symposium}, 
  title={Z{CCloud}: Exploring Wasted Green Power for High-Performance Computing}, 
  year={2016},
  volume={},
  number={},
  pages={1051-1060},
  doi={10.1109/IPDPS.2016.96},
  publisher={IEEE},
  address={New York, NY, USA},
}

@misc{sukprasert2023quantifying,
      title={Quantifying the Benefits of Carbon-Aware Temporal and Spatial Workload Shifting in the Cloud}, 
      author={Thanathorn Sukprasert and Abel Souza and Noman Bashir and David Irwin and Prashant Shenoy},
      year={2023},
      eprint={2306.06502},
      archivePrefix={arXiv},
      primaryClass={cs.DC}
}

@ARTICLE{liu-green-balancing,
  author={Liu, Zhenhua and Lin, Minghong and Wierman, Adam and Low, Steven and Andrew, Lachlan L. H.},
  journal={IEEE/ACM Transactions on Networking}, 
  title={Greening Geographical Load Balancing}, 
  year={2015},
  volume={23},
  number={2},
  pages={657-671},
  doi={10.1109/TNET.2014.2308295}
}

@online{google-dc-shifting,
  author = "Ana Radovanovic",
  year = "2020",
  title = "Our data centers now work harder when the sun shines and wind blows",
  url =  "https://blog.google/inside-google/infrastructure/data-centers-work-harder-sun-shines-wind-blows/",
  month = "April",
  lastaccessed = "February 26, 2024",
  organization={Google},
}

@online{amazon-sustainability-model,
 author = "AWS",
 year = "2023",
 title = "{Sustainability Pillar - AWS Well-Architected Framework}",
  url = "https://docs.aws.amazon.com/wellarchitected/latest/sustainability-pillar/sustainability-pillar.html",
  month = "October",
  lastaccessed = "February 26, 2024"
}

@misc{amazon-sustainability,
  key       = {sustainability-report-aws},
  title     = {2022 {Amazon Sustainability Report}},
  note      = {https://sustainability.aboutamazon.com/2022-sustainability-report.pdf},
  year      = 2022,
  organization = {Amazon}
}

@misc{google-sustainability,
  key       = {sustainability-report-aws-google},
  title     = {{Google Environmental Report 2023}},
  note      = {https://www.gstatic.com/gumdrop/sustainability/google-2023-environmental-report.pdf},
  year      = 2023
}

@ARTICLE{green500,
  author={Feng, Wu-chun and Cameron, Kirk},
  journal={Computer}, 
  title={The {G}reen500 list: Encouraging sustainable supercomputing}, 
  year={2007},
  volume={40},
  number={12},
  pages={50-55},
  doi={10.1109/MC.2007.445}
}

@InProceedings{geopm,
author="Eastep, Jonathan
and Sylvester, Steve
and Cantalupo, Christopher
and Geltz, Brad
and Ardanaz, Federico
and Al-Rawi, Asma
and Livingston, Kelly
and Keceli, Fuat
and Maiterth, Matthias
and Jana, Siddhartha",
editorx="Kunkel, Julian M.
and Yokota, Rio
and Balaji, Pavan
and Keyes, David",
title="Global Extensible Open Power Manager: A Vehicle for {HPC} Community Collaboration on Co-Designed Energy Management Solutions",
booktitle="High Performance Computing",
year="2017",
publisher="Springer International Publishing",
address="Cham",
pages="394--412",
isbn="978-3-319-58667-0"
}

@inproceedings{baolin-sustainable-hpc,
author = {Li, Baolin and Basu Roy, Rohan and Wang, Daniel and Samsi, Siddharth and Gadepally, Vijay and Tiwari, Devesh},
title = {Toward Sustainable {HPC}: Carbon Footprint Estimation and Environmental Implications of {HPC} Systems},
year = {2023},
isbn = {9798400701092},
publisher = {Association for Computing Machinery},
address = {New York, NY, USA},
url = {https://doi.org/10.1145/3581784.3607035},
doi = {10.1145/3581784.3607035},
booktitle = {International Conference for High Performance Computing, Networking, Storage and Analysis},
articleno = {19},
numpages = {15},
location = {<conf-loc>, <city>Denver</city>, <state>CO</state>, <country>USA</country>, </conf-loc>},
series = {SC '23}
}

@INPROCEEDINGS{georgiou-scheduler-incentive,
  author={Georgiou, Yiannis and Glesser, David and Rzadca, Krzysztof and Trystram, Denis},
  booktitle={15th IEEE/ACM International Symposium on Cluster, Cloud and Grid Computing}, 
  title={A Scheduler-Level Incentive Mechanism for Energy Efficiency in {HPC}}, 
  year={2015},
  volume={},
  number={},
  pages={617-626},
  doi={10.1109/CCGrid.2015.101},
  publisher={IEEE},
  address = {New York, NY, USA},
}

@ARTICLE{pham-predicting-cloud,
  author={Pham, Thanh-Phuong and Durillo, Juan J. and Fahringer, Thomas},
  journal={IEEE Transactions on Cloud Computing}, 
  title={Predicting Workflow Task Execution Time in the Cloud Using A Two-Stage Machine Learning Approach}, 
  year={2020},
  volume={8},
  number={1},
  pages={256-268},
  doi={10.1109/TCC.2017.2732344}
}

@misc{slurm-energy-acct,
    title = {{SLURM} Energy Accounting and External Sensors Plug-In},
    author={Auble, Danny and Cadeau, Thomas and  Georgiou, Yiannis and Perry, Martin},
    year={2013},
    note={\url{https://slurm.schedmd.com/SUG13/energy_sensors.pdf}}
}

@INPROCEEDINGS{func-energy-andre,
  author={Schmitt, Norbert and Iffländer, Lukas and Bauer, André and Kounev, Samuel},
  booktitle={IEEE International Conference on Autonomic Computing }, 
  title={Online power consumption estimation for functions in cloud applications}, 
  year={2019},
  volume={},
  number={},
  pages={63-72},
  doi={10.1109/ICAC.2019.00018},
  publisher={IEEE},
  address={Umea, Sweden}
}

@INPROCEEDINGS{smart-watts,
  author={Fieni, Guillaume and Rouvoy, Romain and Seinturier, Lionel},
  booktitle={20th IEEE/ACM International Symposium on Cluster, Cloud and Internet Computing}, 
  title={Smart{W}atts: Self-calibrating software-defined power meter for containers}, 
  year={2020},
  volume={},
  number={},
  pages={479-488},
  doi={10.1109/CCGrid49817.2020.00-45},
  publisher={IEEE},
  address = {New York, NY, USA},
  location={Melbourne, Australia}
}

@inproceedings{boerner-access,
author = {Boerner, Timothy J. and Deems, Stephen and Furlani, Thomas R. and Knuth, Shelley L. and Towns, John},
title = {{ACCESS}: Advancing innovation: {NSF}’s advanced cyberinfrastructure coordination ecosystem: Services \& support},
year = {2023},
isbn = {9781450399852},
publisher = {Association for Computing Machinery},
address = {New York, NY, USA},
url = {https://doi.org/10.1145/3569951.3597559},
doi = {10.1145/3569951.3597559},
booktitle = {Practice and Experience in Advanced Research Computing},
pages = {173–176},
numpages = {4},
location = {Portland, OR, USA},
series = {PEARC '23}
}

@incollection{keahey-chameleon, 
    title={Lessons Learned from the {C}hameleon Testbed},
    author={Kate Keahey and Jason Anderson and Zhuo Zhen and Pierre Riteau and Paul Ruth and Dan Stanzione and Mert Cevik and Jacob Colleran and Haryadi S. Gunawi and Cody Hammock and Joe Mambretti and Alexander Barnes and Fran\c{c}ois Halbach and Alex Rocha and Joe Stubbs}, 
    booktitle={USENIX Annual Technical Conference}, 
    publisher={USENIX Association},
    address={Boston, MA, USA},
    month={July},
    year={2020},
}

@misc{electricity-maps,
    key = {Electricity-Maps},
    title = {Electricity Maps},
    note={\url{https://app.electricitymaps.com/map}},
    year = 2022
}

@article{di2021energy,
  title={Energy-based Accounting Model for Heterogeneous Supercomputers},
  author={Di Pietrantonio, Cristian and Harris, Christopher and Cytowski, Maciej},
  journal={arXiv preprint arXiv:2110.09987},
  year={2021}
}

@inproceedings{chadha2023greencourier,
  title={Green{C}ourier: Carbon-Aware Scheduling for Serverless Functions},
  author={Chadha, Mohak and Subramanian, Thandayuthapani and Arima, Eishi and Gerndt, Michael and Schulz, Martin and Abboud, Osama},
  booktitle={9th International Workshop on Serverless Computing},
  pages={18--23},
  year={2023}
}

@article{sharma2023challenges,
  title={Challenges and opportunities in sustainable serverless computing},
  author={Sharma, Prateek},
  journal={ACM SIGENERGY Energy Informatics Review},
  volume={3},
  number={3},
  pages={53--58},
  year={2023},
  publisher={ACM New York, NY, USA}
}

@article{patros2021toward,
  title={Toward sustainable serverless computing},
  author={Patros, Panos and Spillner, Josef and Papadopoulos, Alessandro V and Varghese, Blesson and Rana, Omer and Dustdar, Schahram},
  journal={IEEE Internet Computing},
  volume={25},
  number={6},
  pages={42--50},
  year={2021},
  publisher={IEEE}
}

@inproceedings{lange-sert,
author = {Lange, Klaus-Dieter and Tricker, Michael G.},
title = {The design and development of the server efficiency rating tool ({SERT})},
year = {2011},
isbn = {9781450305198},
publisher = {Association for Computing Machinery},
address = {New York, NY, USA},
url = {https://doi.org/10.1145/1958746.1958769},
doi = {10.1145/1958746.1958769},
booktitle = {2nd ACM/SPEC International Conference on Performance Engineering},
pages = {145–150},
numpages = {6},
location = {Karlsruhe, Germany},
series = {ICPE '11}
}

@manual{irs-depreciation,
    key = {How To Depreciate
Property},
    title = {How To
Depreciate
Property},
    note={\url{https://www.irs.gov/pub/irs-pdf/p946.pdf}},
    year={2023},
    publisher={{Department of the Treasury Internal Revenue Service}},
    organization={{Department of the Treasury Internal Revenue Service}}
}

@article{kahn-rapl-2018,
author = {Khan, Kashif Nizam and Hirki, Mikael and Niemi, Tapio and Nurminen, Jukka K. and Ou, Zhonghong},
title = {{RAPL} in action: Experiences in Using {RAPL} for power measurements},
year = {2018},
issue_date = {June 2018},
publisher = {Association for Computing Machinery},
address = {New York, NY, USA},
volume = {3},
number = {2},
issn = {2376-3639},
url = {https://doi.org/10.1145/3177754},
doi = {10.1145/3177754},
journal = {ACM Transactions on Modeling and Performance Evaluation of Computing Systems},
month = {mar},
articleno = {9},
numpages = {26}
}

@inproceedings{david2010rapl,
  title={{RAPL}: Memory power estimation and capping},
  author={David, Howard and Gorbatov, Eugene and Hanebutte, Ulf R and Khanna, Rahul and Le, Christian},
  booktitle={16th ACM/IEEE International Symposium on Low Power Electronics and Design},
  pages={189--194},
  year={2010},
  publisher={IEEE},
  address = {New York, NY, USA},
}

@inproceedings{chard-funcx,
author = {Chard, Ryan and Babuji, Yadu and Li, Zhuozhao and Skluzacek, Tyler and Woodard, Anna and Blaiszik, Ben and Foster, Ian and Chard, Kyle},
title = {Func{X}: A federated function serving fabric for science},
year = {2020},
isbn = {9781450370523},
publisher = {Association for Computing Machinery},
address = {New York, NY, USA},
url = {https://doi.org/10.1145/3369583.3392683},
doi = {10.1145/3369583.3392683},
booktitle = {29th International Symposium on High-Performance Parallel and Distributed Computing},
pages = {65–76},
numpages = {12},
location = {Stockholm, Sweden},
series = {HPDC '20}
}

@INPROCEEDINGS{copik-rfaas,
  author={Copik, Marcin and Taranov, Konstantin and Calotoiu, Alexandru and Hoefler, Torsten},
  booktitle={IEEE International Parallel and Distributed Processing Symposium}, 
  title={r{FaaS}: Enabling High Performance Serverless with {RDMA} and Leases}, 
  year={2023},
  volume={},
  number={},
  pages={897-907},
  doi={10.1109/IPDPS54959.2023.00094},
  publisher={IEEE},
  address = {New York, NY, USA},
}

@inproceedings{gonthier-simulator,
  TITLE = {Data-Driven Locality-Aware Batch Scheduling},
  AUTHOR = {Gonthier, Maxime and Larsson, Elisabeth and Marchal, Loris and Nettelblad, Carl and Thibault, Samuel},
  URL = {https://inria.hal.science/hal-04500281},
  BOOKTITLE = {26th Workshop on Advances in Parallel and Distributed Computational Models},
  ORGANIZATIONx = {{38th IEEE International Parallel and Distributed Processing Symposium}},
  YEAR = {2024},
  MONTH = May,
  HAL_ID = {hal-04500281},
  HAL_VERSION = {v1},
  publisher={IEEE},
  address = {New York, NY, USA},
}

@manual{faust,
    author = {{Solem, Ask and Goel, Vineet}},
    title = {{Faust User Manual}},
    year = {2019},
    note = {https://faust.readthedocs.io/en/latest/},
    organization = {Robinhood Markets}
}

@inproceedings{copik2021sebs,
author = {Copik, Marcin and Kwasniewski, Grzegorz and Besta, Maciej and Podstawski, Michal and Hoefler, Torsten},
title = {Se{BS}: A Serverless Benchmark Suite for Function-as-a-Service Computing},
year = {2021},
isbn = {9781450385343},
publisher = {Association for Computing Machinery},
address = {New York, NY, USA},
url = {https://doi.org/10.1145/3464298.3476133},
doi = {10.1145/3464298.3476133},
booktitle = {22nd International Middleware Conference},
pages = {64–78},
numpages = {15},
location = {Qu\'{e}bec city, Canada},
series = {Middleware '21}
}

@misc{ward2021design,
  author = {Ward, Logan},
  title = {{ML}-in-the-loop molecular design with {P}arsl},
  howpublished = {\url{https://github.com/ExaWorks/molecular-design-parsl-demo/tree/main}},
  note = {Accessed: 2024-01-24},
  year={2021}
}

@article{hinz2018cost,
  title={A cost model for {IaaS} clouds based on virtual machine energy consumption},
  author={Hinz, Mauro and Koslovski, Guilherme Piegas and Miers, Charles C and Pilla, La{\'e}rcio L and Pillon, Maur{\'\i}cio A},
  journal={Journal of Grid Computing},
  volume={16},
  pages={493--512},
  year={2018},
  publisher={Springer}
}

@INPROCEEDINGS{kurpicz2016evape,
  author={Kurpicz, Mascha and Orgerie, Anne-Cécile and Sobe, Anita},
  booktitle={24th Euromicro International Conference on Parallel, Distributed, and Network-Based Processing}, 
  title={How Much Does a {VM} Cost? {E}nergy-Proportional Accounting in {VM}-Based Environments}, 
  year={2016},
  volume={},
  number={},
  pages={651-658},
  doi={10.1109/PDP.2016.70},
  publisher={IEEE},
  address={New York, NY, USA}
}

@INPROCEEDINGS{aldossary2018energycost,
  author={Aldossary, Mohammad and Djemame, Karim},
  booktitle={5th International Symposium on Innovation in Information and Communication Technology}, 
  title={Energy-based Cost Model of Virtual Machines in a Cloud Environment}, 
  year={2018},
  volume={},
  number={},
  pages={1-8},
  doi={10.1109/ISIICT.2018.8613288},
  publisher = {IEEE},
  address = {New York, NY, USA},
}

@ARTICLE{narayan2014metering,
  author={Narayan, Akshay and Rao, Shrisha},
  journal={IEEE Transactions on Services Computing}, 
  title={Power-Aware Cloud Metering}, 
  year={2014},
  volume={7},
  number={3},
  pages={440-451},
  doi={10.1109/TSC.2013.22}
}

@inproceedings{margery2017co2,
author = {Margery, David and Guyon, David and Orgerie, Anne-Cecile and Morin, Christine and Francis, Gareth and Palansuriya, Charaka and Kavoussanakis, Kostas},
title = {A {CO2} Emissions Accounting Framework with Market-based Incentives for Cloud Infrastructures},
year = {2017},
isbn = {9789897582417},
publisher = {SCITEPRESS - Science and Technology Publications, Lda},
address = {Setubal, PRT},
url = {https://doi.org/10.5220/0006356502990304},
doi = {10.5220/0006356502990304},
booktitle = {6th International Conference on Smart Cities and Green ICT Systems},
pages = {299–304},
numpages = {6},
location = {Porto, Portugal},
series = {SMARTGREENS 2017}
}

@misc{hoefler-greengraph500,
  author={Torsten Hoefler},
  title={{The Green Graph500}},
  year={2012},
  month={July},
  location={Hamburg, Germany},
  note={\url{http://www.unixer.de/~htor/publications/}},
}

@inproceedings{you2023zeus,
    author = {Jie You and Jae-Won Chung and Mosharaf Chowdhury},
    title = {Zeus: Understanding and Optimizing {GPU} Energy Consumption of {DNN} Training},
    booktitle = {20th USENIX Symposium on Networked Systems Design and Implementation (NSDI 23)},
    year = {2023},
    isbn = {978-1-939133-33-5},
    address = {Boston, MA},
    pages = {119--139},
    url = {https://www.usenix.org/conference/nsdi23/presentation/you},
    publisher = {USENIX Association},
    month = apr
}

@misc{chung2024perseus,
      title={Perseus: Reducing Energy Bloat in Large Model Training}, 
      author={Jae-Won Chung and Yile Gu and Insu Jang and Luoxi Meng and Nikhil Bansal and Mosharaf Chowdhury},
      year={2024},
      eprint={2312.06902},
      archivePrefix={arXiv},
      primaryClass={cs.LG},
      url={https://arxiv.org/abs/2312.06902}, 
}

@inproceedings{wiesner2021letswait,
author = {Wiesner, Philipp and Behnke, Ilja and Scheinert, Dominik and Gontarska, Kordian and Thamsen, Lauritz},
title = {Let's wait awhile: How temporal workload shifting can reduce carbon emissions in the cloud},
year = {2021},
isbn = {9781450385343},
publisher = {Association for Computing Machinery},
address = {New York, NY, USA},
url = {https://doi.org/10.1145/3464298.3493399},
doi = {10.1145/3464298.3493399},
booktitle = {22nd International Middleware Conference},
pages = {260–272},
numpages = {13},
location = {Qu\'{e}bec city, Canada},
series = {Middleware '21}
}

@inproceedings{lyu2023myths,
author = {Lyu, Jialun and Wang, Jaylen and Frost, Kali and Zhang, Chaojie and Irvene, Celine and Choukse, Esha and Fonseca, Rodrigo and Bianchini, Ricardo and Kazhamiaka, Fiodar and Berger, Daniel S.},
title = {Myths and Misconceptions Around Reducing Carbon Embedded in Cloud Platforms},
year = {2023},
isbn = {9798400702426},
publisher = {Association for Computing Machinery},
address = {New York, NY, USA},
url = {https://doi.org/10.1145/3604930.3605717},
doi = {10.1145/3604930.3605717},
booktitle = {2nd Workshop on Sustainable Computer Systems},
articleno = {7},
numpages = {7},
location = {Boston, MA, USA},
series = {HotCarbon '23}
}

@inproceedings {ji24scarif,
  author={Ji, Shixin and Yang, Zhuoping and Cahoon, Stephen and Jones, Alex K and Zhou, Peipei},
  booktitle={IEEE Computer Society Annual Symposium on VLSI}, 
  title={{SCARIF}: Towards Carbon Modeling of Cloud Servers with Accelerators}, 
  year={2024},
  volume={},
  number={},
  pages={1-6},
  doi={},
  publisher={IEEE},
  address={New York, NY, USA},
}

@article{serenari2024greenwhisk,
  title={{G}reen{W}hisk: Emission-aware computing for serverless platform},
  author={Serenari, Jayden and Sreekumar, Sreekanth and Zhao, Kaiwen and Sarkar, Saurabh and Lee, Stephen},
  journal={arXiv preprint arXiv:2409.03029},
  year={2024}
}

@misc{jiang2024ecolifecarbonawareserverlessfunction,
      title={{EcoLife}: Carbon-Aware Serverless Function Scheduling for Sustainable Computing}, 
      author={Yankai Jiang and Rohan Basu Roy and Baolin Li and Devesh Tiwari},
      year={2024},
      eprint={2409.02085},
      archivePrefix={arXiv},
      primaryClass={cs.DC},
      url={https://arxiv.org/abs/2409.02085}, 
}

@article{lin2024bridging,
  title={Bridging the Sustainability Gap in Serverless through Observability and Carbon-Aware Pricing},
  author={Lin, Changyuan and Shahrad, Mohammad},
  journal={HotCarbon'24},
  year={2024}
}

@misc{passmark,
    key = {{P}ass{M}ark {CPU} Benchmark},
    title = {{CPU} Benchmarks},
    year={2024},
    author={{PassMark}},
    url = {https://www.cpubenchmark.net/},
    address = {Surry Hills, NSW, Austrailia}
}

@misc{lumi2024sustainability,
    key={{LUMI}},
    title={{LUMI}: Sustainable Future},
    year={2024},
    url={https://www.lumi-supercomputer.eu/sustainable-future/}
}

@article{kamatar2024greenfaas,
  title={GreenFaaS: Maximizing Energy Efficiency of HPC Workloads with FaaS},
  author={Kamatar, Alok and Hayot-Sasson, Valerie and Babuji, Yadu and Bauer, Andre and Rattihalli, Gourav and Hogade, Ninad and Milojicic, Dejan and Chard, Kyle and Foster, Ian},
  journal={arXiv preprint arXiv:2406.17710},
  year={2024}
}

@inproceedings{tirmaze2020borg,
author = {Tirmazi, Muhammad and Barker, Adam and Deng, Nan and Haque, Md E. and Qin, Zhijing Gene and Hand, Steven and Harchol-Balter, Mor and Wilkes, John},
title = {Borg: the next generation},
year = {2020},
isbn = {9781450368827},
publisher = {Association for Computing Machinery},
address = {New York, NY, USA},
url = {https://doi.org/10.1145/3342195.3387517},
doi = {10.1145/3342195.3387517},
booktitle = {Proceedings of the Fifteenth European Conference on Computer Systems},
articleno = {30},
numpages = {14},
location = {Heraklion, Greece},
series = {EuroSys '20}
}

@misc{sci,
    key={Software Carbon Intensity ({SCI}) Specification},
    title = {Software Carbon Intensity ({SCI}) Specification},
    year = {2024},
    url={https://sci.greensoftware.foundation/},
    publisher={Green Software Foundation}
}

@INPROCEEDINGS{solorzano2024toward,
  author={Solórzano, Ana Luisa Veroneze and Sato, Kento and Yamamoto, Keiji and Shoji, Fumiyoshi and Brandt, Jim M. and Schwaller, Benjamin and Walton, Sara Petra and Green, Jennifer and Tiwari, Devesh},
  booktitle={SC24: International Conference for High Performance Computing, Networking, Storage and Analysis}, 
  title={Toward Sustainable HPC: In-Production Deployment of Incentive-Based Power Efficiency Mechanism on the Fugaku Supercomputer}, 
  year={2024},
  volume={},
  number={},
  pages={1-16},
  doi={10.1109/SC41406.2024.00030},
  publisher={IEEE},
  address={New York, NY, USA},
}

@inproceedings{bashir2024sunkcarbon,
author = {Bashir, Noman and Gohil, Varun and Subramanya, Anagha Belavadi and Shahrad, Mohammad and Irwin, David and Olivetti, Elsa and Delimitrou, Christina},
title = {The Sunk Carbon Fallacy: Rethinking Carbon Footprint Metrics for Effective Carbon-Aware Scheduling},
year = {2024},
isbn = {9798400712869},
publisher = {Association for Computing Machinery},
address = {New York, NY, USA},
url = {https://doi.org/10.1145/3698038.3698542},
doi = {10.1145/3698038.3698542},
booktitle = {Proceedings of the 2024 ACM Symposium on Cloud Computing},
pages = {542–551},
numpages = {10},
location = {Redmond, WA, USA},
series = {SoCC '24}
}

@INPROCEEDINGS{kishwar2018contractbased,
  author={Ahmed, Kishwar and Bull, Jesse and Liu, Jason},
  booktitle={2018 IEEE Intl Conf on Parallel and Distributed Processing with Applications, Ubiquitous Computing and Communications, Big Data and Cloud Computing, Social Computing and Networking, Sustainable Computing and Communications (ISPA/IUCC/BDCloud/SocialCom/SustainCom)}, 
  title={Contract-Based Demand Response Model for High Performance Computing Systems}, 
  year={2018},
  volume={},
  number={},
  pages={580-589},
  doi={10.1109/BDCloud.2018.00091},
  publisher = {IEEE},
  address = {New York, NY, USA},
}

@inproceedings{pauloski2024taps,
    author = {Pauloski, J. Gregory and Hayot-Sasson, Valerie and Gonthier, Maxime and Hudson, Nathaniel and Pan, Haochen and Zhou, Sicheng and Foster, Ian and Chard, Kyle},
    title = {{TaPS: A Performance Evaluation Suite for Task-based Execution Frameworks}},
    address = {New York, NY, USA},
    booktitle = {IEEE 20th International Conference on e-Science},
    doi = {10.1109/e-Science62913.2024.10678702},
    pages = {1-10},
    publisher = {IEEE},
    year = {2024}
}

@inproceedings{roy24hiddencarbon,
author = {Roy, Rohan Basu and Kanakagiri, Raghavendra and Jiang, Yankai and Tiwari, Devesh},
title = {The Hidden Carbon Footprint of Serverless Computing},
year = {2024},
isbn = {9798400712869},
publisher = {Association for Computing Machinery},
address = {New York, NY, USA},
url = {https://doi.org/10.1145/3698038.3698546},
doi = {10.1145/3698038.3698546},
booktitle = {Proceedings of the 2024 ACM Symposium on Cloud Computing},
pages = {570–579},
numpages = {10},
location = {Redmond, WA, USA},
series = {SoCC '24}
}

@inproceedings{sharma2024accountable,
author = {Sharma, Prateek and Fuerst, Alexander},
title = {Accountable Carbon Footprints and Energy Profiling For Serverless Functions},
year = {2024},
isbn = {9798400712869},
publisher = {Association for Computing Machinery},
address = {New York, NY, USA},
url = {https://doi.org/10.1145/3698038.3698531},
doi = {10.1145/3698038.3698531},
booktitle = {Proceedings of the 2024 ACM Symposium on Cloud Computing},
pages = {522–541},
numpages = {20},
location = {Redmond, WA, USA},
series = {SoCC '24}
}

\end{document}